\documentstyle[aps,twocolumn,prb,epsf]{revtex}

\begin{document}

\twocolumn[
\hsize\textwidth\columnwidth\hsize\csname@twocolumnfalse\endcsname
\draft
\title{{\bf Fundamental Limit on ``Interaction Free'' Measurements}}
\author{Steven H. Simon and P. M. Platzman \\
Lucent Technologies Bell Labs, Murray Hill, NJ, 07974}
\date{\today}

\maketitle

\begin{abstract}
  In ``interaction free'' measurements, one typically wants to detect
  the presence of an object without touching it with even a single
  photon.  One often imagines a bomb whose trigger is an extremely
  sensitive measuring device whose presence we would like to detect
  without triggering it.  We point out that all such measuring devices
  have a maximum sensitivity set by the uncertainty principle, and
  thus can only determine whether a measurement is ``interaction
  free'' to within a finite minimum resolution. We further discuss
  exactly what can be achieved with the proposed ``interaction free''
  measurement schemes.
\end{abstract}

]

In a highly influential recent paper by Elitzur and Vaidman\cite{EV},
it was pointed out that the presence of an object (often called a
``bomb'') can often be discerned without it absorbing even a single
photon.  This ``interaction free measurement'' scheme and later
improvements on it\cite{Kwiat,Serious,EnergyExchange,Neutrons} have
received a lot of attention, both in the popular press\cite{Popular}
as well as in serious scientific journals
\cite{Kwiat,Serious,EnergyExchange,Neutrons,DV}.  In this paper we
wish to re-examine such measurement schemes and consider how they may
be limited by the Heisenberg uncertainty principle.

We would like to be very precise about what we mean by an
``interaction free'' measurement, and we attempt to define this in
terms of a specific bomb detection experiment.  We imagine that the
bomb we wish to detect has a trigger that is so sensitive that it will
explode if interacts in any way with any particles that are sent to
probe it -- I.e., if it scatters or absorbs any of these particles.
This bomb trigger should be sensitive to an arbitrarily small momentum
transfer from the probe particle to the bomb, as well as being
sensitive to angular momentum transfer, energy transfer, and transfer
of any other quantum number we could consider.  We now imagine that
some gnome challenges us to determine if he/she has placed this
sensitive bomb within some predetermined region (denoted by the dotted
box\cite{Endnote2} in Fig.  1).  If we succeed in detecting the
presence of this bomb without blowing it up, we will have performed an
``interaction free'' measurement.  We note, however, that the
measurement can only be declared to be ``interaction free'' if the
bomb is truly an ideal detector.  If the bomb trigger is unreliable,
then we will never know if we have interacted with the bomb or not
(This will become important below).

Performing an interaction free measurement as defined above may seem
impossible at first --- and indeed, within classical physics such a
thing would clearly be forbidden. However, by exploiting wave-particle
duality, a number of groups have suggested
\cite{Kwiat,Serious,EnergyExchange,Neutrons} that such measurements
are in fact possible.  Below, we will discuss the simplest of these
proposed measurement schemes, and our results will apply more
generally.  In this paper we will point out that these schemes in fact
do not satisfy the definition of ``interaction free'' given above.  We
then continue on to ask ourselves what precisely is achieved by
these schemes.  In particular, we will show that schemes can indeed
claim to be ``energy exchange free'' (as first discussed in Ref.
\onlinecite{EnergyExchange}) or free from transfer of certain other
quantum numbers, but are not free from transfer of all quantum
numbers.  Specifically, we will show that such experiments are not
free of momentum transfer (although they can be made to have
``minimal'' momentum transfer).

\begin{figure}[htbp]
\begin{center}
  \leavevmode \epsfbox{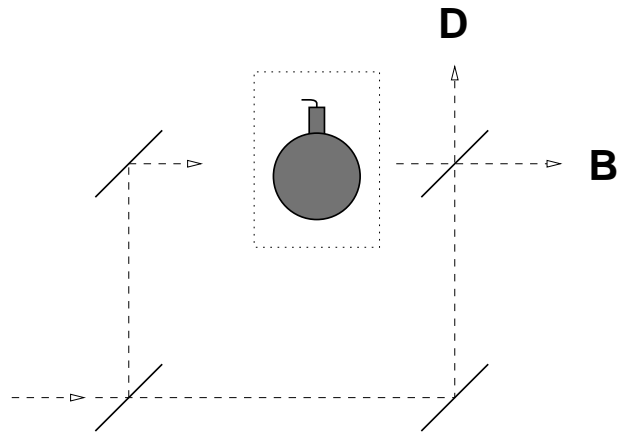}
\end{center}
\caption{The Mach-Zehnder Interferometer.  Beam Splitters have
  reflectivity of 50\%.  When the beam-line is clear, the interference
  is arranged such that all of the incoming light exits toward
  detector B (bright) and none of it exits toward detector D (dark).
  When the upper beam-line is blocked by an object, 50\% of the
  incoming photons are absorbed by the object, 25\% of the incoming
  photons exit towards detector B, and 25\% exit towards detector D.
  Thus, if we do the experiment with a single photon, and if we happen
  to detect that photon at detector D, then we know that the object is
  blocking the beam-line even though the object has not absorbed a
  single photon.}
\end{figure}

We begin by discussing the simplest so-called ``interaction-free''
measurement scheme.  As described above, we will think in terms of a
bomb detection experiment.  The scheme for detecting the bomb,
originally proposed by Elitzur and Vaidman\cite{EV}, is to construct a
Mach-Zehnder interferometer as shown in Fig. 1.  We arrange the length
of the arms of the interferometer to be such that the interference is
constructive when a photon exits towards detector B (for bright) and
destructive when it exits toward detector D (for dark).  Thus, so long
as the beam lines are not blocked by any objects, all of the light
that enters the interferometer exits towards detector B.

Now we consider what happens when the gnome places the bomb in the
predetermined region\cite{Endnote2} (I.e., in the dotted box in Fig 1)
such that the bomb blocks the beam-line and prevents interference of
the two paths of light.  For the moment, let us assume that the bomb
is in some sense a perfect absorber --- an assumption that we will see
below has some difficulties.  With this assumption, when the bomb is
blocking the beam line, 50\% of the light sent into the interferometer
will be absorbed by the object, 25\% will exit towards detector B, and
25\% will exit towards detector D.  (We have also assumed here that
our beam splitters have a reflectivity of 50\%.).  We then send a
single photon into the interferometer.  50\% of the time this photon
will be absorbed by the bomb and it will explode.  However, 25\% of
the time, we will detect the photon at detector D, which is normally
dark, and we will know that the bomb is blocking the beam-line without
it having absorbed the photon (Also 25\% of the time the photon comes
out at detector B which is inconclusive).  Thus, in this simple way,
we are able to perform what appears to be an ``interaction free''
measurement at least some fraction of the time\cite{Kwiat,Endnote1}.
Experiments of this type have indeed been performed\cite{Kwiat,DV} (in
one case with single photons\cite{Kwiat}), albeit with imperfect
detectors and with a ``bomb trigger'' with finite sensitivity.

What we would like to point out in this paper is that there is a
fundamental limit on the possible sensitivity of the bomb, and hence
the measurement can only be considered ``interaction free'' to within
this limited sensitivity. 

In order to understand the source of this limitation, we consider the
preparation of the experiment.  In order for the gnome to set up the
experiment and place the bomb in the pre-arranged
region\cite{Endnote2} (the dotted box in Fig. 1), he/she must know the
position of the bomb to within some uncertainty $\Delta x$.  Since
there is now a finite uncertainty of position, the bomb must have a
momentum uncertainty of $\Delta p = \hbar/\Delta x$.  If the bomb were
sensitive to momentum changes this small, then it would be triggered
by quantum fluctuations (and would therefore be a useless device).
Another way to say this is that the gnome would be unable to put the
sensitive bomb in place without triggering it.  

To make this important point more explicit, we imagine how the trigger
of such a bomb might work.  Before we do our experiment, the gnome
places the bomb in the prearranged region\cite{Endnote2} (I.e., in the
dotted box) in some wave-packet such that $\Delta x$ is known
sufficiently well for the gnome to know that the bomb is indeed in
this region.  After we shoot our photon though the apparatus, the
trigger apparatus measures the momentum of the bomb.  If
the momentum is sufficiently large, then the gnome knows that we must
have transferred momentum to the bomb (and the gnome would then make
the bomb explode).  However, the initial momentum state of the bomb
must have an uncertainty of $\hbar/ \Delta x$, so the gnome certainly
cannot reliably detect if we transfer any momentum less than this
amount to the bomb.   It is interesting to note that this fundamental
limit arises from understanding the measuring device (the bomb
trigger) as a quantum mechanical device itself. 

Because of this limit on the sensitivity of the bomb, it is clear that
that no measurement can ever be ``interaction free'' by the definition
given above (I.e., the bomb detection experiment with an infinitely
sensitive bomb trigger as defined in the second paragraph of this
paper), since any bomb can always recoil a very small amount and this
interaction could not be detected.  One might object that the reason
no experiment fits our above definition is simply because our
definition is overly restrictive. This may indeed be the case.
(Although we also note that the experiment described above seems a
reasonably natural choice in the absence of any prior attempts at a
definition).  Although our choice of definition is a matter of
nomenclature which should not overly concern us, it remains a physically
meaningful question to ask ``what {\it can} be achieved by these
so-called interaction free measurement schemes?.''  

It is clear that in order to actually conduct an experiment similar to
that proposed above, we must concede that the bomb will have a
sensitivity limit for momentum transfers (although it may remain
arbitrarily sensitive to transfers of other quantum numbers).  Let us
then consider an experiment analogous to that described above, but
conducted with a finitely sensitive bomb such that only momentum
transfers larger than the order of $\hbar/\Delta x$ will cause it to
explode.  This modified bomb is now sufficiently insensitive so as not
to be triggered by the quantum fluctuations of momentum which are
necessarily present due to the uncertainty principle.  With such a
modified bomb of finite sensitivity, we should not declare that detection
of this bomb is truly ``interaction free'' since we will never know if
the bomb has interacted very weakly with a probe particle.
Nonetheless, it is certainly true that the above described
interferometric measurement scheme\cite{EV} (as well as more
sophisticated versions of interferometric
schemes\cite{Kwiat,Serious,EnergyExchange,Neutrons}) can indeed detect
the presence of this modified bomb without blowing it up.  We might
say that this is now a ``minimum interaction'' measurement (By which
we mean, we can detect a maximally sensitive bomb without triggering
it).

It is now interesting to ask if there are other, perhaps simpler,
methods of detecting this modified --- slightly less sensitive --
bomb without blowing it up.  (I.e., of performing a similar ``minimum
interaction'' measurement).  One would only need to arrange to touch
the bomb extremely softly to detect its presence, and so long as the
transferred momentum remains less than $\hbar/\Delta x$, the bomb will
not blow up.

One might guess that we could simply probe such a bomb with very long
wavelength photons (or other probe particles), thus using a momentum
transfer below the bomb's sensitivity limit.  One must be careful,
however, being that the bomb may still be sensitive to other quantum
numbers of the probe particles -- such as energy or angular momentum,
and the bomb might still explode if it absorbs the long wavelength
photon even though the {\it momentum} transfer is below the
sensitivity limit.  In other words, we have pointed out above that the
bomb cannot be arbitrarily sensitive to momentum transfers (and we
have agreed to make our bomb only finitely sensitive to momentum) but
the bomb may still remain arbitrarily sensitive to other properties of
the probe particle.  Thus in order to perform a ``minimum
interaction'' measurement, we must arrange that no other quantum
numbers of the probe particle are changed in the course of the
interaction.

One particularly simple approach to making such ``minimum
interaction'' measurements is to perform a simple small angle
scattering experiment.  We imagine sending a plane wave of short
wavelength light at the bomb (Here, the beam must be a wide enough
wave packet to be able to either hit the bomb or diffract around the
bomb).  For a bomb, assumed to be a perfect absorber, the absorption
cross section is on the order of the cross sectional area of the
object\cite{Jackson}.  However, there is also an elastic scattering
cross section for small angle diffraction around the edge of the
object (I.e., shadow scattering) that is also on the order of the
cross sectional area of the object\cite{Jackson} (with factors that
depend on the precise geometry and boundary conditions).  The angle of
the diffraction is typically on the order of $1/(k_{in} a)$ where
$k_{in}$ is the wavevector ($k=2 \pi/\lambda$) of the incident light
and $a$ is the length scale of the object.  The momentum transfer to
the object when a single incident photon of momentum $p_{in} = \hbar
k_{in}$ makes one of these small angle scattering events is then given
by (roughly) $p_{in}/(k_{in}a ) = \hbar/a$.  In our experiment the
length of the sample $a$ is also the necessary uncertainty in the
position $\Delta x$ (since we must know the position to within a
distance $a$ to make sure the object blocks the beam-line).  Thus, the
momentum transfer $p = \hbar/\Delta x$ in a small angle elastic
scattering event is a ``minimal interaction''.  We see that by
performing a simple scattering experiment with short wavelength single
photons we can perform such a ``minimal interaction'' measurement
which is in many ways equivalent to the interference scheme discussed
above.  Here, we send single short wavelength photons (in a plane wave
state) at the bomb, and measure the outgoing momentum of the photon.
In some fraction of trials, the photon comes out with the same
momentum as it went in, which tells us nothing (analogous to measuring
a photon in detector B above).  In some fraction of trials the photon
is either absorbed, or is elastically scattered by a large angle, in
which case the bomb blows up.  However, in some fraction of the
trials, we measure that the photon undergoes small angle scattering,
and we have discerned the presence of the bomb without triggering it.

As a final note, we consider a slight variant of this scattering
experiment.  Here, we imagine holding the bomb in a very weak harmonic
potential well to localize its position.  The bomb, being itself a
quantum mechanical object, is placed in the ground state wavefunction
of the harmonic potential.  Again, because its position is known to
within some accuracy $\Delta x$, it has a momentum uncertainty $\Delta
p = \hbar/\Delta x$.  We note that the (harmonic oscillator) energy
levels of the bomb in the well are discrete and are spaced by an
energy of order $\Delta E = (\Delta p)^2/(2 M)$ where $M$ is the mass
of the bomb.  If we try to transfer some small momentum less than
$\Delta p$ to the bomb (either by using long wavelength photons or
small angle scattering), we would not be able to give the bomb enough
energy to reach the next eigenstate of the harmonic well.  Therefore,
the bomb must remain in the ground state wavefunction and the momentum
would be transferred directly to the well itself.  Indeed, measuring
the excitation state of the bomb in the well is a maximal sensitivity
measurement since it can measure momentum transfers of order $\Delta p
= \hbar/\Delta x$ and one could never have a bomb trigger more
sensitive than this.

In summary, we have pointed out that all measuring devices have a
maximum sensitivity fixed by the uncertainty principle.    One can then
always perform an ``interaction free measurement'' (in the sense of
determining the presence of the bomb without triggering it) by simply
probing very softly with very low momentum transfer (either small
angle scattering or long wavelength photons).  We believe that a large
range of so called ``interaction free'' schemes may
have similar limitations once the quantum mechanical nature of the
measuring devices are properly understood. 

The authors acknowledge helpful conversations with M. Andrews,
L. Balents, D. Morin, and O. Narayan.


\begin{references}
\bibitem{Endnote2} Most generally, the prearranged region (the dotted
  box) might be of any given size.  It is clear, however, that for the
  interferometric scheme of Ref. \onlinecite{EV} to work, this
  prearranged region must be small enough such that the bomb will
  always block the beam-line if placed within this region.  In this
  case, we must have the region on the order of the size of the bomb
  in the transverse direction, and on the order of the length of the
  beam line in the longitudinal direction.
\bibitem{Endnote1} The simple Mach-Zehnder interferometer is in some
  sense a rather inefficient experiment, being that it blows up the
  bomb quite frequently.  We can define an efficiency as $P_d/(P_e +
  P_d)$ where $P_d$ is the probability of detecting the bomb and $P_e$
  is the probability of exploding the bomb\cite{Kwiat}.  It has been
  pointed out that by using a multiple reflection geometry equivalent
  to a resonant cavity\cite{Kwiat}, one can make the experiment
  virtually 100\% efficient ---in the sense that we almost never blow
  up the bomb. For our current purposes, however, the simple
  interferometer will be a sufficient example, and the results will
  apply more generally.
\end{references}
\end{document}